main paper  latex file
fig one     latex file
(There are 15 postscript figure files which makes this
submission too long.  If you wish the figures, please
send a mail message and I will email them to you.

Richard Haymaker   haymaker@rouge.phys.lsu.edu)

\documentstyle[preprint,revtex]{aps}
\tightenlines
\begin{document}
\draft
\preprint{LSUHE 94-159}
\def\overlay#1#2{\setbox0=\hbox{#1}\setbox1=\hbox to \wd0{\hss #2\hss}#1%
\hskip -2\wd0\copy1}
\begin{title}
Distribution of the color fields around static quarks: \\
Flux tube profiles
\end{title}
\author{ Richard~W.~Haymaker, ~Vandana~Singh, and~Yingcai~Peng }
\begin{instit}
Department of Physics and Astronomy, \\
Louisiana State University, Baton Rouge, Louisiana 70803-4001
\end{instit}
\author{Jacek Wosiek}
\begin{instit}
Jagellonian University, Institute of Physics, \\
Reymonta 4 Cracow, Poland; \\
and Max-Planck-Institut f\"{u}r Physik - Werner-Heisenberg-Institute \\
P.O. Box 40 12 12 \\
Munich, Germany
\end{instit}
\begin{abstract}
We report detailed calculations of the profiles of energy and action
densities in the quark-antiquark string in SU(2) lattice gauge theory.
\end{abstract}
\narrowtext
\section{Introduction}
\label{sec: intro}

Although gross features of the string connecting a quark and an antiquark
are understood, little is known about the detailed structure of the flux
tube.  Questions remain as to the size and shape of the energy distribution.
The linearly rising potential can be understood if the energy per unit length
of the string is independent of the separation of the quark and antiquark.
That still allows for a variation of the peak density and the width.  We are
also interested in the fields themselves, for example whether they are
governed by dual superconductivity, whether they appear coulomb like
at small separations as expected from asymptotic freedom, and what
excited strings look like.

This paper addresses the issue of the energy distribution at zero temperature.
 The difficulties
of getting good signal-to-noise measurements at large $q \overline{q} $
separations are well
known.  There are trade-off's in every strategy.  Our choice is to represent
the $q \overline{q}$ sources via a Wilson loop.  We then attempt to extrapolate
our results to large time extent of the loops.  The alternative is to use
Polyakov lines to represent the sources.  But then the 'area law' forces one
to work at finite temperature and then extrapolate to zero temperature.

There have been some analytical predictions of flux tube properties.
L\"{u}scher, Munster, and Weisz have studied the
bosonic string model of the effective tube~\cite{luscher1}. Their results
show that the flux tube width has a logarithmic behavior with the
quark separation $R$. Adler constructed a dielectric model~\cite{adler},
which predicts that the energy peak density in the flux tube
$\sim \frac{1}{R}$, and the flux tube width $\sim \sqrt{R}$. Baker,
Ball and Zachariasen have obtained the flux tube solution for QCD
from the dual formulation of the Yang-Mills theory. They give analytical
evidence for flux tube formation in QCD~\cite{baker}.

In this paper we report detailed calculations of
the spacial profiles of energy and action distributions of the flux tube.
Sec.~\ref{sec: sim} pulls together some of the simulation details for
completeness.

There are six orientations of plaquettes corresponding to the six
components of chromoelectric and chromomagnetic fields.  We assign the
plaquette value to the field-squared at the center of the plaquette.
Hence all components are defined at different space-time points.
Interpolation in four dimensions is required in order to add these
functions to come up with the energy density.  The most obvious method
such as cubic splines in four dimensions is not very satisfactory and
tends to give rather obvious interpolation artifacts in surface plots. We
decided on a method used by commercial surface plot software termed
'Kriging.'  It adapted nicely to four dimensions without biasing any
particular direction in four-space.  This is described in
Sec.~\ref{sec: com}.  In this way we can fill in values at every
point on a lattice with lattice constant a/2.

Also in this section we discuss the parametrization of the flux tube
profile on the plane perpendicular to the line joining the quark and
antiquark at the midpoint between them.  It is very difficult to get
accurate moments of these profiles since they are dominated by values
of the flux at large transverse distance where data is very
scant.  Our compromise was to find a good fitting function with a very
economical parametrization.  We settled on a simple function that is
exponential for large distance and gaussian for small distances.
It did remarkably well in that the fit was very insensitive to the
sample of points used to fit over a very wide range of transverse
distances.

Having determined the profile parameters for a wide range of Wilson
loops, we then extrapolate these parameters to large time extent of
the Wilson loops.   In Sec.~\ref{sec: gsr} we carry out this
extrapolation and present results.

\section{Simulation}
\label{sec: sim}

The lattice observable needed to measure the flux is
the following~\cite{fn,fo,sommer1,sommer2,wh1}.
\begin{eqnarray}
    f^{\mu \nu}(x) &=& \frac{\beta}{a^4} \left(
\frac{\langle W P^{\mu \nu}_x \rangle }{\langle W \rangle}
-  \langle P \rangle \right),
\nonumber\\
 &\approx&  \frac{\beta}{a^4} \left(
\frac{\langle W P^{\mu \nu}_x - W P^{\mu \nu}_{x_{R}} \rangle}
{\langle W \rangle}  \right),
\label{e30}
\end{eqnarray}
where $W$ is the Wilson loop, $P^{\mu \nu}_x$ the plaquette
located at $x$, $\beta = \frac{4}{g^2}$ and $x_{R}$ is a distant
reference point.  In the classical continuum limit
\begin{equation}
  f^{\mu \nu} \stackrel{a \rightarrow 0} {\longrightarrow} -\frac{1}
 {2} \langle ( F^{\mu \nu} )^{2} \rangle_{q\overline{q} - vac},
\label{e40}
\end{equation}
where the notation $ \langle \cdots \rangle_{q\overline{q}-vac} $ means
the difference
of the average values in the $q\overline{q}$ and vacuum state. From now
on we shall be using field components in Minkowski space notation 1and hence
\begin{equation}
  f^{\mu \nu} \rightarrow
   \frac{1}{2} ( -B_1^2,-B_2^2,-B_3^2; E_1^2,E_2^2,E_3^2).
\label{e50}
\end{equation}
Correspondence between various components and $f^{\mu \nu}$ is standard:
space-space plaquettes are magnetic,  space-time plaquettes are electric.
The energy and action densities are respectively
\begin{eqnarray}
\epsilon = \frac{1}{2} ( E^2 + B^2),
\nonumber\\
\gamma   = \frac{1}{2} ( E^2 - B^2).
\label{e60}
\end{eqnarray}
Since the magnetic contribution turns out to be negative, there is
a strong cancellation between the two terms in the energy, whereas they
are enhanced in the action. One remark is that the measured negative
magnetic energy density $B^2$ was obtained from the difference between
the value of the $q\bar q$ state and that of the vacuum, as shown
in Eq.~(\ref{e40}).
To understand the physical origin of the negative $B^2$, one need
to study the dynamical properties of $SU(N)$ vacuum, which has not
been fully understood.

In the earlier paper~\cite{hw} we gave some of the details of
the simulation and measurement techniques.  Briefly, our lattice size is
$ 17^3 \times 20 $.  We ran three values of $\beta=$ 2.3, 2.4 and 2.5.
The corresponding lattics spacings $a(\beta)$ were taken to be
a(2.3) = 0.171 fm, a(2.4) = 0.128 fm, and a(2.5) = 0.089 fm.  We measured
Wilson loops for all sizes up to $7 \times 9$ which are reported in
Ref.~\cite{hw} as well as the eigenvalues of the transfer matrix calculated
from these loops.

An important ingredient in our simulation is the analytic multihit
technique in which one evaluates link integrals of the operators
being measured.  When this is applied to a single link on a Wilson loop
the result can be given in closed form.
\begin{equation}
\int [dU] U e^{\frac{\beta}{2} tr[UK^{\dagger}]} =
\frac{I_2 (\beta b)}{I_1 (\beta b)} \; V \;
\int [dU]e^{\frac{\beta}{2} tr[UK^{\dagger}]},
\label{2.1}
\end{equation}
where K is the sum of six 'staples' coupling to given U in the action
\cite{ppr}, and $I_n(x)$ is the modified Bessel function.
The sum of $SU(2)$ matrices in the $ j= \frac{1}{2} $ representation
is a multiple of an $SU(2)$ matrix which we denote by V
\begin{equation}
K = b V  ; \;\; b \equiv   ( det K )^{\frac{1}{2}},
\label{2.2}
\end{equation}
Effectively what has happened is that the link has been replaced by a
sum of staples associated with that link times a gauge invariant weight
given by the ratio of Modified Bessel functions.

Fig.~\ref{f10} shows the links involved.  The original links in the
Wilson loop and plaquette are the bold lines.  The staple links
are also shown.  It is clear from this picture that not all the links
on the Wilson loop can be evaluated with Eqn.~(\ref{2.1}).  The criterion
to be able to use Eqn.~(\ref{2.1}) to evaluate a particular (bold)
link is that
the corresponding staples do not involve any of the other bold links.
A straight line of links satisfies this criterion, but links forming
a corner do not.
Therefore the plaquette and the corners of the Wilson loop
have to be treated separately.  Further
since we are trying to measure the self energy of the
quarks we need to bring the plaquette close to and in fact touching the
Wilson loop which will result in further special cases.
It is expedient to measure the Wilson loop and plaquette
for each position and orientation independently and then evaluate the
cross correlation.   This is fine as long as a bold link on the
Wilson loop does not overlap with a staple link on the multihit plaquette
or vice versa.  Since we make use of fast fourier transforms to do the
cross correlations all relative  positions of the two are automatically
obtained.  We must then go back and recalculate the correct exceptions
separately.  We have the choice of either doing more integrals analycially
or drop back to fewer analytic integrations for these exceptions.  We chose
the latter, maintaining all possible analytic integrations that
could be constructed with the above three analytic results.

Our compromise in all these special cases was the following:  We calculated
(a) the plaquette and multihit plaquette, (b) the Wilson loop with
all multihit links and the Wilson loop with a gap as shown in
Fig.~\ref{f10}. Then the
correlations were done using three cases: (i) For large separations of the
Wilson loop and plaquette, all links were multihit. (ii) As the plaquette
is brought close to the Wilson loop, the multihit plaquette
was dropped.  (iii) When the plaquette was brought in contact with
the Wilson loop, certain multihit links were also dropped as indicated by the
gap in Fig.~\ref{f10}.  This compromise means that
having chosen the gap in one particular direction
we can not get self energy data at other places on the loop for example to
explore the self energy for loop with T and R reversed.

The plaquette and corner integrals can not be done in closed form
but can be evaluated using a character expansion.
Details can be found in the appendix of Ref.~\cite{hw2}.  For completeness
we would like to pull together just the results.

\subsection{ Character Expansions}
\label{subsec: cha}

The basic technique used here is to expand integrands using group characters
as basis functions. The character is given by the trace of the
$(2 j + 1)$ dimensional rotation matrices.
\begin{equation}
\chi^{(j)}(U) = \sum_{m} D^{(j)}_{m m}(U) = \frac{\sin{((2 j + 1)\psi)}}
{\sin{\psi}}; \;\;\; j = 0, \frac{1}{2}, 1, \frac{3}{2}, \cdots .
\label{a1}
\end{equation}
We parametrize group elements by an axis of rotation $\hat{n}$
and an angle of rotation about
that axis denoted here by  $2 \psi$.
\begin{equation}
U = \cos{(\psi)} + i \sin{(\psi)} \hat{n} \cdot \vec{\tau},
\label{a2}
\end{equation}
and the group manifold is then the hypersurface of the 4-sphere and the
invariant group integration measure is uniform on this manifold.

We will use the character expansion of a link or product of links in
the action:
\begin{equation}
e^{\frac{\beta}{2} tr(U)} = \sum_j c_j(\beta)
\chi^{(j)} (U); \;\;\;
c_j(\beta)=(4j+2) \frac{I_{2j+1}(\beta)}{\beta},
\label{a4.2}
\end{equation}
where $I_{2j+1}(\beta)$ is the modified Bessel function.  The sum is over
all representations as indicated in Eq.~(\ref{a1}).
All integrals can be done making use of the orthogonality of the
group integration,
\begin{equation}
\int [dU] D^{(j)}_{m n}(U)^{*} D^{(j')}_{m' n'}(U)
= \frac{2 \pi^2}{2j+1}\delta_{j j'}\delta_{m m'}\delta_{n n'}.
\label{a4.2a}
\end{equation}
{}From this relation we can find for example,
\begin{equation}
\int [dU] \chi^{(j)} (VU) \chi^{(j')} (U)
= \frac{2 \pi^2}{2j+1}\delta_{j j'} \chi^{(j)}(V),
\label{a4.2b}
\end{equation}
where $V$ is also a link or product of links.
This formula is very useful in evaluating the following integrals.

\subsection{Corner Integral}
\label{subsec: cor}

Consider the integral over two links that form a corner
\begin{eqnarray}
\langle U_{2} U_{1} \rangle \equiv \frac{1}{Z} \int [dU_{1} dU_{2}]
U_{2} U_{1} \; e^{-S},
\nonumber \\
Z \equiv \int [dU_{1} dU_{2}]  \; e^{-S},
\label{a20}
\end{eqnarray}
where the relevent terms in the action are
\begin{equation}
-S \equiv \frac{\beta}{2}
(tr(W^{\dagger}U_{2}U_{1}) + b_{1} tr(U_{1} V_{1}^{\dagger})
+ b_{2} tr(U_{2} V_{2}^{\dagger})).
\label{a21}
\end{equation}
Z can be evaluated immediately using Eq.~(\ref{a4.2b}),
\begin{equation}
Z = \sum_j \frac{(2 \pi^2)^2}{(2j+1)^{2}} c_{j}(\beta) c_{j}(\beta b_1)
c_{j}(\beta b_2)
\chi^{(j)} (P).
\label{a22}
\end{equation}
where
\begin{equation}
P \equiv  W^{\dagger} V_{2} V_{1}.
\label{a23}
\end{equation}
Applying the character expansion to the three terms in the action we give
the final result:
\begin{eqnarray}
&&\langle U_2 U_1 \rangle = \frac{1}{Z} \sum_j \frac{(2 \pi^2)^2}{(2j+1)^2}
c_j(\beta b_1) c_j(\beta b_2) \{ W c_j^{\prime} \chi^{(j)}(P) +
\nonumber \\
&&  c_j(\beta)\frac{ [V_2 V_1 - W \frac{1}{2} \chi^{(\frac{1}{2})}(P)]
[(j+1) \chi^{(j - \frac{1}{2})}(P) - j \chi^{(j + \frac{1}{2})}(P)]}
{\beta (1 - \frac{1}{4} \chi^{(\frac{1}{2})}(P)^2)} \}
\label{a33}
\end{eqnarray}
where $c_j^{\prime} = d c_j(\beta)/d \beta$.

\subsection{Plaquette Integral}
\label{subsec: pla}

In the plaquette integral, at most one link can be evaluated with
Eq.~(\ref{2.1}).  Further it is clear that only two of the links
forming a corner can make use of Eq.~(\ref{a33}).  Since the
plaquette is a gauge invariant trace of now four links, this case
is in fact considerably simpler than the corner integral.  Again the
details can be found in Ref.~\cite{hw2}.
Consider the following four-link integral
\begin{equation}
Z(\gamma, \beta) \equiv \int [dU_1 dU_2 dU_3 dU_4] \; e^{-S},
\label{a12}
\end{equation}
where
\begin{equation}
-S \equiv (\frac{\gamma}{2}
tr[U_4^{\dagger} U_3^{\dagger} U_2 U_1] + \frac{\beta}{2}
\sum_{k=1}^4 tr[U_k K_k^{\dagger}]).
\label{a12.1}
\end{equation}

We have displayed explicitly the dependence of the action on the
links making up a particular plaquette.  Each $K_k$ matrix is the sum of
{\it five} staples.  The results are
\begin{equation}
Z(\gamma, \beta) = \sum_j c_j(\gamma) \frac{(2 \pi^2)^4}{(2j+1)^4}
\prod_{k=1}^4 c_j(\beta b_j) \chi^{(j)}(V_4^{\dagger} V_3^{\dagger} V_2 V_1)
\label{a16}
\end{equation}
Pulling this together, we get the result
\begin{equation}
\int [ \prod_l dU_l] \frac{1}{2}
tr(U_4^{\dagger} U_3^{\dagger} U_2 U_1)\; e^{-S} =
\frac{ \frac{\partial Z(\gamma,\beta)}{\partial \gamma}\vert_{\gamma = \beta}}
{Z(\beta,\beta)}
\int [ \prod_l dU_l] \; e^{-S} .
\label{a17}
\end{equation}

  Using the plaquette and corner integrals one can reduce
statistical fluctuations in the simulations significantly,
because the measured values of plaquette and corner of
Wilson loop involve many more links.

\section{Data Interpolation and Fitting}
\label{sec: com}

     The raw flux data can be calculated from Eq.~(\ref{e30})
by Monte Carlo simulations on the lattice.
To obtain a smooth flux distribution and extract physical
parameters, we used the following methods to interpolate
the raw data and fit the raw data with the profile function.

\subsection{Interpolation}
\label{subsec: inte}

The six components of the flux $f^{\mu \nu}(x)$
are defined at the center of the
corresponding plaquette and hence are all at different space-time
points. Finding a good interpolation method in four dimensions is
not as simple as it might seem.
In order to combine them we employed a four dimensional
interpolation method that minimizes the variance of the estimated
value of the function.  In other words as one varies the value of
each input point over say one standard deviation, one requires
that the linear interpolated function be
as insensitive as possible.  There is much literature on this under the
name of Kriging~\cite{david} in mining engineering and a two dimensional
version is employed in surface plot software.

Consider a function, $f(x)$, that is given as a statistical variable at a set
of
points, $x_j$.  Suppose the mean, $\langle f(x_j) \rangle$,
and variance
$\sigma_{x_{j}}^2 = \langle(f(x_j) - \langle f(x_j) \rangle)^2\rangle$ are
known.
Define an estimate of the function, $\tilde{f}(x)$, as the linear combination
\begin{equation}
\tilde{f}(x) = \sum_j a_j f(x_j).
\label{3.10}
\end{equation}
where the sum is over a subset of the points, and the coefficients
$a_j$'s are determined by the set of points $\{x_j \} $ and the point
$x$. Now use this estimate to evaluate the function at one of the
given points $x_i$ and
consider the variance of the difference of the function $f(x_i)$ and its
estimate $\tilde{f}(x_i)$
\begin{eqnarray}
\sigma_e^2 &=& \langle ((f(x_i)-\langle f(x_i) \rangle)
 - ( \tilde{f}(x_i) -\langle \tilde{f}(x_i)\rangle) )^2 \rangle
\nonumber\\
&=& \sigma_{x_i}^2 - 2 \sum_j a_j \sigma_{x_i,x_j}^2
+  \sum_{j j'} a_j a_{j'} \sigma_{x_j,x_{j'}}^2.
\label{3.20}
\end{eqnarray}
The sum excludes the point $x_i$.  The covariance matrix is defined
\begin{equation}
 \sigma_{x_k,x_l}^2 =
\langle (f(x_k)-\langle f(x_k) \rangle)(f(x_l)-\langle f(x_l) \rangle) \rangle.
\label{3.30}
\end{equation}

The Kriging method determines the coefficients $a_j$ by minimizing
$\sigma_e^2$  subject to a constraint, i.e. we minimize
\begin{equation}
{\cal L}(a_j) =   \sigma_e^2(a_j) + 2 \mu(\sum_j a_j - 1),
\label{3.40}
\end{equation}
where $\mu$ is a lagrange multiplier.  The constraint arises from
the condition that  when applied to all points in the set,
$\tilde{f}(x_i)$ underestimates $f(x_i)$ as often as it overestimates it:
\begin{equation}
\langle\langle \sum_j a_j f(x_j) -  f(x_i) \rangle\rangle = 0.
\label{3.50}
\end{equation}
Using $\langle \langle f(x_i) \rangle \rangle = f$
this gives the constraint $\sum_j a_j = 1$. Hence the coefficients are
given by the equations
\begin{eqnarray}
\sum_{j'} \sigma_{x_j,x_{j'}}^2 a_{j'} + \mu &=& \sigma_{x_i,x_j}^2,
\nonumber\\
\sum_{j'} a_{j'} = 1.
\label{3.60}
\end{eqnarray}

We want to apply this to interpolate $f(x)$ at a general point.
The $a_j$'s are determined from the covariance matrix
at the general point. Since the $a_j$'s are independent
of the overall normalization of the covariance matrix and the
relevant property is how the covariance matrix varies with distance
between the two arguments,  so in practical
applications one assumes that the covariance matrix
falls as a function of distance, going to zero at large separations.

Our choice is:
\begin{equation}
\sigma_{x_j,x_{j'}}^2 \sim \exp{(-|j-j'|/\tau}).
\label{3.70}
\end{equation}
One can easily check that
if one chose the set $\{x_j\}$ to include the point $x_i$, then the
corresponding $a_i=1 $, and all other $a_j$'s equal zero. For a point
near one of the $x_j$, the corresponding $a_j=1 $ dominates.
Since we have an ad hoc ansatz for the covariance matrix the actual
errors in our data do not enter in the interpolation.  In fact this
is just a method of linear interpolation which weighs the nearby points
more than the far points which is divorced from its statistical origins.

The freedom in this method is in the choice of neighboring points
and the value of the correlation length $\tau$.  Our criterion
was to get a smooth interpolation which did not produce variations
of the order of a lattice spacing.  We found $\tau \sim $ the lattice
spacing and a radius that included one or two dozen points in the four
dimensional neighborhood of the interpolated point.
The errors were calculated by linearly interpolating the errors
from the neighboring points.
This overestimates the errors since it assumes
that the values used to interpolate are completely correlated.
Assuming they are completely uncorrelated then interpolating the variance
gives an error typically less than half of the linearly interpolated
errors.  Since we do not know the covariance matrix we took the worst
case.

Using this method we filled in every point at the half lattice spacing
for each component which then allowed us to combine any components.
Figs.~\ref{f20} and ~\ref{f30} give surface plots of the interpolated
data for a Wilson loop of $T=7$ and $R=5$ with $\beta=2.4$.  Shown
are two sections, one through the two quarks and the second on a plane
midway between the two quarks to show the flux tube profile.

Fig.~\ref{f20} shows the electric and magnetic components
$\frac{1}{2}E_{\|}^2$, $-\frac{1}{2}B_{\|}^2$,
$\frac{1}{2}E_{\perp}^2$, and  $-\frac{1}{2}B_{\perp}^2$.
{}From these plots we can see that
only the electric components have prominent peaks around the quarks,
and the two transverse components, $\frac{1}{2}E_{\perp}^2$
 and  $-\frac{1}{2}B_{\perp}^2$, are
essentially the same.  These two components cancel in calculating
the energy and add in the action. They are the widest of the four
profiles and hence the action has a wider profile than the energy.

Fig.~\ref{f30} shows the action and energy for the same parameters
as above. The peaks around the quarks
rise above the background equally in the energy and action since they
come predominantly from the electric components. The Michael sum
rules confirm this behavior as described in the earlier
paper~\cite{hw}. The flux tube profiles for action and
energy are quite different.  Note the large cancellation in the
energy compared to the action and the resulting difficulty in
measuring the former.  Rotational symmetry is reasonably well
restored in the action profiles but there are large lattice
artifacts in the energy profiles.

\subsection{Fit to Profile Function}
\label{subsec: fit}

The basic parameters we are interested in is to find the peak value
and widths of the flux density of the energy and action.  Obtaining
good data on the second moment of the profile in order to determine
its width is problematic.  The profiles fall sharply and the fourth
moment needed to estimate error would be sampling only large distances
from the axis where the signal to noise is small.  Instead we
fitted the energy and action density in the plane at the midpoint
between $q$ and $\bar{q}$ using the function
\begin{eqnarray}
f(r_{\perp}) = a \exp\Bigl({-\sqrt{c^2 + (r_{\perp}/b)^2}}\Bigr).
\label{e70a}
\end{eqnarray}
or rewriting in terms of the width at half maximum, $B$
\begin{eqnarray}
f(r_{\perp}) = A e^C \exp\Bigl({-\sqrt{C^2 + (2 C ln2 +
ln^22)(r_{\perp}/B)^2}}\Bigr).
\label{e70b}
\end{eqnarray}

The peak value, $A$, and the width $B$,  were very well
determined
using a $\chi^2$ fit for each of 70 cases of different loop
sizes and values of $\beta$. (We used Minuit to determine the fit and
errors.)
For the third parameter we chose the decay length of
the tail of this function, $D= \sqrt{(2 C ln2 + ln^22)}/B$.
This parameter was less well determined. We attempted to include an additive
constant to the function since our sum rule paper gave evidence for it.
However this gave too much freedom to the fit and gave unstable answers.
We also tried a gaussian plus background constant and this did not fit
nearly as well as the function chosen. In conclusion, the data is not
good enough to determine four parameters, and the best function we found
has some ambiguity in the choice but the peak value, $A$, and width
at half maximum $B$ are very well determined.  The exponential
fall off of this function is perhaps modeling a true exponential
falloff plus possibly a background which we believe is there due to
the fact that we are measuring the flux relative to a reference point
that is not at infinity.

 Our complete fitting results are shown in Tables~\ref{t10}-\ref{t60}
  for action and energy distributions.
%
%
The four numbers for
each loop size are the peak value, $A$ in $GeV/fm^3$, the width at half
maximum,
$B$ in $fm$, the exponential decay length, $D$ in $fm$, and the integration
of the area under the curve to give the string tension in $GeV/fm$.
As a reminder, since we are looking at the middle time slice and looking
at the mid-point of the flux tube, these parameters are measured
in the  $x,y$ plane for $z,t$ at the middle of the Wilson loop.

Our characterization of a good fit requires explanation.  We do not
report the $\chi^2$ because the fit is used to average out the
lattice artifacts in the rotational symmetry.  For small loops the
errors in the data are quite small but rotational symmetry is not
equally good.  As a consequence $\chi^2$ can be very large
for fits since the fitting function is rotationally invariant.  When
the errors in the data approach the size of the lattice artifacts
$\chi^2$/degree of freedom drops to the order 1.    We believe the
fit is good because of the stability of the numbers over changing
the set of points used in the fit.  We determined the parameters
using 9 sets of points, choosing the interior of circles of radius $1.5a$
to $5.5a$. (Recall our interpolation produces values at every point
at half a lattice spacing.) For all our fits the variation in the
parameters over this set of points was of the order or less that the
statistical Minuit errors quoted for the parameters in the tables.
We did not fold in this systematic error. However we believe we have
overestimated errors in the interpolation and we believe the
effects of these two contributions are comparable.

Fig.~\ref{f40} illustrates a general feature of our data.
The cluster of three points
for each $R$ and $T$ correspond to the three quantities:
\begin{eqnarray}
\epsilon &=&
 \frac{1}{2} (E_{\|}^2 + B_{\|}^2)
         + \frac{1}{2} (E_{\perp}^2 + B_{\perp}^2),
\nonumber\\
\epsilon_{\|} &=& \frac{1}{2} ( E_{\|}^2 + B_{\|}^2),
\nonumber\\
\epsilon (T \leftrightarrow R)&=& \frac{1}{2} ( E_{\|}^2 + B_{\|}^2)
         - \frac{1}{2} (E_{\perp}^2 + B_{\perp}^2).
\label{e80}
\end{eqnarray}
If one turns the
Wilson loop on its side, the $\|$ components are unchanged but the
$\perp$ components of the electric
and magnetic fields are reversed:
$E_{\perp}^2 \leftrightarrow -B_{\perp}^2$. Hence there is a sign change
in the third expression.
The central points of the cluster are the $\|$ components only.
{\it The clustering of the points implies that the $\perp$ components
of the electric and magnetic contributions to energy density
are approximately equal but of opposite sign and cancel.}
The width of the peak is even less sensitive to the transverse components
giving essentially the same value for all three points.

\section{Results for The Ground State }
\label{sec: gsr}

\subsection{Extrapolation to Infinite Wilson Loop Time Extent}
\label{subsec: ext}

In the companion paper~\cite{hw} we described the use of the trasfer
matrix eigenvalues to extrapolate the data to infinite time.
Similarly to the potential case, the finite time extent of the Wilson
loop $W(R,T)$ introduces contamination of the $f^{\mu \nu}(x)$ by the
excited states of the colour field~\cite{hpsw1}.  From the transfer matrix
representation of the correlation between the Wilson loop $W$ and
a plaquette $P$,
\begin{equation}
\langle W P \rangle = Z^{-1} Tr ({\cal T}^{L_t - T/a} S
   {\cal T}_{q\bar q}^{T/2a} P {\cal T}_{q\bar q}^{T/2a} S),
\label{3.1}
\end{equation}
where $S$ is the operator which excites the $q\bar q$ states from
the vacuum, and ${\cal T}_{q\bar q}$ is the transfer matrix
in the $q \bar q$ sector.
We obtain for any component $f^{\mu \nu}$ $(\equiv F)$
\begin{equation}
F = {\cal F} + {\cal R}_1 e^{-E_1 T} + {\cal C}_1 e^{-E_1 T/2} + \cdots
\label{3.2}
\end{equation}
where ${\cal F}$ denotes true ground state average, and $E_1$ stands for
the energy gap of ${\cal T}_{q\bar q}$,
 which can be determined by fitting the
Wilson loops to the exponentials as described in Ref.~\cite{hw}.
Formula~(\ref{3.2}) applies to any component ($\mu \nu$) and any
three-space location $\vec{x}$ of the plaquette
$P^{\mu \nu}(\vec{x}, x_0)$. To minimize the effect of the
cross terms, $x_0$ was always chosen at the middle of the time
interval of the Wilson loop (hence the term $e^{-E_1 T/2}$).
For any parameter extracted from the
flux distribution $f^{\mu \nu}$, e.g., the peak density, one can
use Eq.~(\ref{3.2}) to fit the parameter values corresponding to
various Wilson loop time extents, and extract the asymptotic value
${\cal F}$ for $T\rightarrow \infty$.

A sample of the results of extrapolating are shown in Figs.~\ref{f50},
{}~\ref{f60} and ~\ref{f70}.  One can see that the
extrapolation is reasonable. However, there are some cases in which
the fit did not work, one could trace the problem to
too large a variation for small T where the points are most accurate.
This indicates that a constant and one exponential is not sufficient
to take for small T. We then dropped points as needed at the small T.
These cases were all for $\beta= 2.5$ which correspond to the
smallest physical times.

\subsection{Discussion of Scaling Plots for Parameters}
\label{subsec: dsp}

   After we extrapolate the measured physical quantities to
infinite Wilson loop time extent, we can study the behaviors of
these quantities with the quark separation $R$.
In Figs.~\ref{f80} and~\ref{f90} we just the results
for the peak densities and flux widthes.

The basic issue is whether the peak value of the energy density
stabilizes to a constant or goes to zero with quark separation.
This is not easy to settle as can be seen in Fig.~\ref{f80} (a).
Roughly speaking
we know from the linearly rising potential that the string tension
$\sim$ (width)$^2$ $\times$ (peak value) should be constant.
Both Figs.~\ref{f80} and~\ref{f90} show that we are marginally asymptotic
in quark separation, $R$.
The two curves in Fig.~\ref{f80} are $\sim 1/R^4$ and $\sim 1/R$.
 A coulomb field
would fall like the former but since the string tension is constant
the asymptotic width would have to grow like $R^2$ which clearly it does not.
Therefore we can rule out a coulomb field as expected.
Interestingly for small separations,
the eyeball fit to  $\sim 1/R^4$ is quite good
which may be due to a coulomb like behavior at small distances.
(The above argument that the width must grow like $R^2$ does not apply
because there is no string for small $R$.)
The dielectric model~\cite{adler} predicts  the peak density $\sim 1/R$.
Such a behavior would imply the width $\sim \sqrt{R}$
which is certainly possible in our data.  We can say that
the peak energy density and width are consistent with
a constant value for large quark separation but we can not rule out
a slow variation.  The issue can be tightened by making use of
the Michael sum rules~\cite{mic} as we mention in the next section.

\subsection{Using Sum Rules to Tighten Predictions of Energy Behavior}
\label{subsec: sru}

A consistency check on the flux distributions can be obtained
by using the Michael sum rules~\cite{mic} for energy and action.
\begin{eqnarray}
 \frac{1}{2} \sum_{\vec{x}} ( E(\vec{x})^2 + B(\vec{x})^2 )&=& E_{0}(R) ,
\nonumber\\
\frac{1}{2}
\sum_{\vec{x}} ( E(\vec{x})^2 - B(\vec{x})^2 ) &=&  -\beta
\frac{\dot{a}}{a} [ E_{0}(R) - \frac{c(\beta)}{a}] -  \beta
\frac{\dot{c}(\beta)}{a}.
\label{e90}
\end{eqnarray}
Here $E_{0}(R)$ is given by
\begin{equation}
\langle W (R, T) \rangle = \sum_i A_i e^{-E_i(R) T},
\label{e91}
\end{equation}
and ($ \cdot \equiv \frac{d}{d\beta}$).  In Ref.~\cite{hw} we have shown that
our data are essentially consistent with these sum rules.  The one difficulty
is the fact that the self energy,  $c(\beta)/a(\beta)$,
determined from the potential differs from the self energy determined
from the the action sum rule.  This may be due to an ambiguity in the
definition of self energy or possibly due to our classical
expressions for energy and action which ignores quantum corrections.
By taking a derivative of these expressions with respect to the
quark separation, R, this difficulty is avoided.
This gives the relation
\begin{equation}
\sigma_A = -\beta\frac{\dot{a}}{a} \sigma;
\;\;\;\;
\sigma_A \equiv
\frac{1}{2}\sum_{\vec{x}_{\perp}} ( E(\vec{x})^2 - B(\vec{x})^2 ); \\
\;\;\;\;
\sigma \equiv
\frac{1}{2}\sum_{\vec{x}_{\perp}} ( E(\vec{x})^2 + B(\vec{x})^2 ).
\label{e100}
\end{equation}
The sums are now over the plane midway between the $q \bar{q}$ pair.

Fig.~\ref{f110} shows the data for the ratio,
$\sigma_A / \sigma$, which were measured for $\beta=2.3$, 2.4 and 2.5,
at various $q\bar q$ separations $R$. One can see that for small
$R$ the ratio is small, however, at large $R$ most data
show large values ($\approx  10. $) for this ratio, but the
fluctuations at large $R$ make an estimate difficult.
{}From the sum rules above and Fig.~\ref{f110} we estimate that the
$\beta$ function, $-\beta \dot{a}/a) \approx  10. \pm 2.$, compared
to the current estimates~\cite{hw,sommer2},  $ 7.7\pm 1.0 $.
The asymptotic value is $-51/121 + 3\pi^2\beta/11
\approx 6.0$ for $\beta=2.4$.  There is ample evidence from other
measurements
that although scaling works well, asymptotic scaling is
violated~\cite{mic92}
and hence we do not expect to get the asymptotic value.

An alternative approach is to assume the sum rules are correct and use
them to infer information about the energy density from the action
density which is far easier to measure  since relative errors
are down by an order of magnitude.  As is clear from the sum rule
the action does not scale yet the variation over these values of $\beta$
is very small.  An examination of Fig.~\ref{f80} (a) shows that the action
for each $\beta$ seems to stabilize to a constant for increasing
distance for the peak density and for the width.
This is strongly supported by data for $\beta =2.3$ and 2.4.
For $\beta = 2.5$ the quark separation
$R$ appears to be too small to draw a conclusion. From these data
one can see that the peak values seem to approach a nonzero finite
constant at large $R$.
We would like to use the sum rules to predict the behavior of the
energy density.  A constant peak energy density follows only
if the widths of the energy and action peaks have the same behavior.
Figure~\ref{f95} shows that in fact they do.
 From this and using the sum rules we conclude that the energy density
stabilizes to a constant value also.
This conclusion is an argument against the dielectric model~\cite{adler}.
However due to the limitation on the available range of $R$ of our data
we can not make conclusions about logarithmic behavior of the
flux tube width as predicted by L\"{u}scher~\cite{lmw1}.

\subsection{Longitudinal Profile of Flux Tube}
\label{subsec: who}

  To see the profile of the whole flux tube, in Fig.~\ref{f150} we plot
the integrations of flux action on various transverse slices along the
$q\bar q$ axis, where $Z=0$ is the location of one quark source, the
positions of $Z<0$ corresponds to slices outside the $q\bar q$ pair,
and the positions with $Z>0$ corresponds to slices between the
$q\bar q$ pair. Because of the symmetry of flux distribution about
the midway point between the $q\bar q$ pair, we only show the
distribution from one end to the midpoint between the $q\bar q$ pair.
One can see that the peak values are
in the neighborhood of the source, which is caused by the large
self-energy concentrated around the source. However, the flux
action values between the $q\bar q$ pair are comparable to the
peak values. This implies that the integrations of flux action on
the transverse slice near the $q\bar q$ pair are not dominated by
the self-energy contributions, although the density values due to
self-energy are large, their distribution regions are very narrow.
This agrees with the action density distribution shown in
Fig.~\ref{f30} (a), where the peaks of action density are restricted
in a narrow region at each source. From Fig.~\ref{f150} one can
also see that away from the outside of the $q\bar q$ sources
the flux action decreases rapidly with the distance from the source,
and almost vanishes at the moderate distance, $R=4a$ (i.e., $Z=-4$).
However, the flux action between the $q\bar q$ pair is large, and
almost keeps to be a constant as $Z$ increases. This agrees with
the expectation for the flux tube formation.

\bigskip
\nonum
\section{Acknowledgments}
 We would like to thank A. Kotanski,
G. Schierholz and T. Suzuki for many fruitful discussions on this
problem.  R.W.H, V.S. and Y.P. are
supported by the DOE under grant DE-FG05-91ER40617. J.W. is supported
in part by the Polish Government under Grants Nos. CPBP 01.01 and CPBP
01.09.

\newpage

\begin{center}

TABLES
\end{center}

\vspace{0.4in}

 TABLE I. The fitting parameters for the action distributions with
$\beta = 2.3$ and various Wilson loop sizes, $T$ and $R$. For each
loop size we list in order the values of the three parameters, $A$ in
Gev/$fm^3$, $B$ in $fm$ and $D$ in $fm$.
The last value in each case gives the integration of the area under
the curve in Gev/$fm$.

\vspace{0.4in}

 TABLE II. The fitting parameters for the energy distributions with
$\beta = 2.3$ and various Wilson loop sizes, $T$ and $R$. Which
is similar to TABLE I.

\vspace{0.4in}

 TABLE III. The fitting parameters for the action distributions with
$\beta = 2.4$ and various Wilson loop sizes, $T$ and $R$. Which
is similar to TABLE I.

\vspace{0.4in}

 TABLE IV. The fitting parameters for the energy distributions with
$\beta = 2.4$ and various Wilson loop sizes, $T$ and $R$. Which
is similar to TABLE I.

\vspace{0.4in}

 TABLE V. The fitting parameters for the action distributions with
$\beta = 2.5$ and various Wilson loop sizes, $T$ and $R$. Which
is similar to TABLE I.

\vspace{0.4in}

 TABLE VI. The fitting parameters for the energy distributions with
$\beta = 2.5$ and various Wilson loop sizes, $T$ and $R$. Which
is similar to TABLE I.

\newpage

\begin{table}
\begin{tabular}{|l|l|l|l|l|l|l|}
    T   & 3 & 4 & 5 & 6 & 7 & 8  \\
    R   &   &   &   &   &   &    \\ \hline
3        &53.0(1)   &47.9(1)   &44.9(2)   &43.4(3)  &40.4(9)  &40.4(9)    \\
         &0.1328(2) &0.1495(4) &0.1597(6) &0.167(1) &0.176(4) &0.176(4)    \\
         &0.1041(2) &0.1105(3) &0.1175(6) &0.121(1) &0.136(4) &0.136(4)    \\
         &6.58(3)   &7.21(5)   &7.69(8)   &8.05(14) &8.58(46) &8.58(46)    \\
\hline
4        &          &41.7(1)   &38.6(2)   &37.7(5)  &34.(1)   &34.(1)    \\
         &          &0.1729(5) &0.189(1)  &0.196(3) &0.198(8) &0.198(8)    \\
         &          &0.1177(4) &0.123(1)  &0.135(2) &0.153(9) &0.153(9)    \\
         &          &7.95(6)   &8.55(14)  &9.2(3)   &9.(1)    &9.(1)    \\
\hline
5        &          &          &34.5(3)   &34.6(7)  &42.(3)   &42.(3)    \\
         &          &          &0.210(3)  &0.213(5) &0.20(2)  &0.20(2)    \\
         &          &          &0.107(1)  &0.128(3) &0.16(2)  &0.16(2)    \\
         &          &          &8.5(2)    &9.3(6)   &12.(3)   &12.(3)    \\
\hline
6        &          &          &          &31.(1)   &39.(4)   &    \\
         &          &          &          &0.22(1)  &0.19(2)  &    \\
         &          &          &          &0.058(3) &0.036(4) &    \\
         &          &          &          &6.7(8)   &7.(1)    &    \\
\end{tabular}
\caption{ }
\label{t10}
\end{table}

\newpage

\begin{table}
\begin{tabular}{|l|l|l|l|l|l|l|}
    T   & 3 & 4 & 5 & 6 & 7 & 8  \\
    R   &   &   &   &   &   &     \\ \hline
3        &24.1(1)   &18.4(1)   &16.0(2)    &14.1(4)  &13.8(6)  &13.(1)    \\
         &0.0813(4) &0.0897(8) &0.0925(15) &0.091(4) &0.096(5) &0.093(9)    \\
         &0.0787(5) &0.0817(9) &0.0878(18) &0.092(4) &0.085(6) &0.088(10)    \\
         &1.33(2)   &1.17(3)   &1.13(5)    &1.02(10) &0.98(14) &0.92(22)    \\
\hline
4        &17.7(1)   &13.2(2)   &10.4(3)    &9.0(6)   &8.(1)    &7.(1)    \\
         &0.094(1)  &0.104(1)  &0.109(4)   &0.115(9) &0.12(2)  &0.12(1)    \\
         &0.088(1)  &0.093(2)  &0.115(6)   &0.112(12)&0.10(2)  &0.020(2)    \\
         &1.28(3)   &1.12(4)   &1.13(11)   &1.0(2)   &0.8(4)   &0.47(14)    \\
\hline
5        &15.6(2)   &10.8(4)   &7.8(4)     &6.(2)    &3.(1)    &    \\
         &0.0924(17)&0.105(4)  &0.12(1)    &0.14(7)  &0.20(5)  &    \\
         &0.097(2)  &0.102(5)  &0.16(3)    &--       &0.024(6) &    \\
         &1.21(6)   &0.99(11)  &1.3(3)     &--       &--       &    \\ \hline
6        &13.2(4)   &9.0(6)    &6.(1)      &5.(2)    &         &    \\
         &0.100(4)  &0.118(9)  &0.13(3)    &0.2(1)   &         &    \\
         &0.106(5)  &0.099(9)  &0.11(3)    &0.04(2)  &         &    \\
         &1.21(12)  &0.93(18)  &--         &--       &         &    \\ \hline
7        &12.3(7)   &8.(2)4    &4.(2)      &         &         &    \\
         &0.095(8)  &0.12(2)   &0.10(3)    &         &         &    \\
         &0.115(14) &0.14(5)   &0.14(8)    &         &         &    \\
         &1.16(23)  &1.2(6)    &--         &         &         &    \\ \hline
8        &10.5(8)   &8.(1)     &           &         &         &    \\
         &0.098(5)  &0.15(2)   &           &         &         &    \\
         &0.137(12) &0.21(7)   &           &         &         &    \\
         &1.2(6)    &--        &           &         &         &    \\
\end{tabular}
\caption{ }
\label{t20}
\end{table}

\newpage

\begin{table}
\begin{tabular}{|l|l|l|l|l|l|l|l|}
    T   & 3 & 4 & 5 & 6 & 7 & 8 & 9 \\
    R   &               &&&&&&        \\ \hline
3        & 103.3(1)  &87.6(1)   &77.7(2)   &72.4(3)   &69.7(4)   &67.1(5)
&65.(1)   \\
         & 0.0947(1) &0.1065(2) &0.1156(3) &0.1218(4) &0.1249(6) &0.128(1)
&0.132(2) \\
         & 0.0796(1) &0.0874(2) &0.0938(3) &0.0970(4) &0.1001(6) &0.102(1)
&0.098(2) \\
         & 6.87(2)   &7.22(3)   &7.45(4)   &7.58(6)   &7.68(9)   &7.76(13)
&7.7(3)   \\  \hline
4        &           &69.42(9)  &58.8(1)   &53.8(2)   &50.3(3)   &49.0(4)
&49.(1)   \\
         &           &0.1262(2) &0.1427(3) &0.1523(6) &0.160(1)  &0.162(1)
&0.165(5) \\
         &           &0.0970(2) &0.1023(3) &0.1066(5) &0.104(1)  &0.111(1)
&0.119(5) \\
         &           &7.60(3)   &7.81(5)   &7.99(7)   &7.9(1)    &8.11(17)
&8.7(6)   \\  \hline
5        &           &          &48.7(2)   &44.6(3)   &40.8(4)   &40.2(7)
&41.(2)   \\
         &           &          &0.1653(6) &0.178(1)  &0.187(2)  &0.186(4)
&0.19(1)  \\
         &           &          &0.1101(5) &0.120(1)  &0.127(2)  &0.151(6)
&0.10(1)  \\
         &           &          &8.22(8)   &8.64(15)  &8.7(3)    &9.2(5)
&8.2(1.2) \\  \hline
6        &           &          &          &39.3(4)   &37.3(8)   &38.(3)
&27.(2)        \\
         &           &          &          &0.195(2)  &0.204(5)  &0.22(2)
&0.28(3)        \\
         &           &          &          &0.121(2)  &0.134(4)  &0.18(3)
&0.034(3)        \\
         &           &          &          &8.7(3)    &9.1(5)    &12.(2)
&9.7(1.7)        \\  \hline
7        &           &          &          &          &32.(2)    &24.(3)   &
    \\
         &           &          &          &          &0.26(1)   &0.23(3)  &
    \\
         &           &          &          &          &0.031(2)  &0.027(3) &
    \\
         &           &          &          &          &9.4(9)    &5.7(1.5) &
    \\
\end{tabular}
\caption{ }
\label{t30}
\end{table}

\begin{table}
\begin{tabular}{|l|l|l|l|l|l|l|l|}
    T   & 3 & 4 & 5 & 6 & 7 & 8 & 9 \\
    R   &           &&&&&&        \\ \hline
3        &50.4(1)   &38.7(2)   &31.5(2)  &28.5(3)  &26.8(4)  &25.4(6)  &25.(2)
   \\
         &0.0623(2) &0.0694(3) &0.073(1) &0.073(1) &0.075(1) &0.073(2)
&0.078(6)   \\
         &0.0599(2) &0.0624(3) &0.067(1) &0.069(1) &0.069(2) &0.073(3)
&0.068(6)   \\
         &1.63(1)   &1.47(2)   &1.33(3)  &1.24(4)  &1.20(6)  &1.17(9)
&1.16(22)   \\  \hline
4        &35.0(2)   &25.0(1)   &17.7(2)  &15.0(3)  &12.9(3)  &12.1(5)  &12.(2)
   \\
         &0.0689(4) &0.0814(5) &0.084(1) &0.092(2) &0.094(3) &0.093(6) &0.09(2)
   \\
         &0.0696(5) &0.0703(5) &0.087(2) &0.084(2) &0.090(4) &0.091(8) &0.05(1)
   \\
         &1.45(2)   &1.26(2)   &1.12(4)  &1.00(6)  &0.93(7)  &0.88(13) &0.6(3)
   \\  \hline
5        &28.4(2)   &18.8(2)   &11.7(2)  &10.2(4)  &8.2(6)   &8.1(8)
&7.8(1.6)   \\
         &0.070(1)  &0.084(2)  &0.097(3) &0.101(6) &0.11(1)  &0.100(9) &0.15(2)
   \\
         &0.076(1)  &0.085(2)  &0.106(5) &0.108(9) &0.14(3)  &0.14(3)  &--
   \\
         &1.31(3)   &1.16(4)   &1.02(7)  &0.94(13) &1.05(26) & --      &0.8(3)
   \\  \hline
6        &23.9(3)   &15.1(3)   &8.6(4)   &7.2(5)   &5.(1)    &5.3(1.1)
&2.5(1.6)   \\
         &0.071(1)  &0.095(2)  &0.108(7) &0.12(1)  &0.14(4)  &--       &--
   \\
         &0.085(2)  &0.087(3)  &0.117(13)&0.11(2)  &0.14(8)  &--       &--
   \\
         &1.26(6)   &1.08(6)   &0.91(14) &0.8(2)   &0.8(5)   &3.1(1.5) &--
   \\  \hline
7        &22.7(3)   &13.8(5)   &7.3(6)   &4.0(7)   &&&        \\
         &0.070(1)  &0.092(5)  &0.107(13)&0.19(4)  &&&        \\
         &0.089(2)  &0.101(8)  &0.13(3)  &0.05(1)  &&&        \\
         &1.23(5)   &1.10(14)  &0.87(25) &0.7(4)   &&&        \\  \hline
8        &20.9(6)   &12.9(7)   &6.(1)    &5.(3)    &&&        \\
         &0.071(2)  &0.096(6)  &0.106(27)&0.06(3)  &&&        \\
         &0.094(5)  &0.105(11) &0.12(5)  &--       &&&        \\
         &1.22(9)   &1.1(2)    &0.7(4)   &--       &&&        \\  \hline
9        &22.(2)    &14.(2)    &6.(2)    &7.(5)    &&&        \\
         &0.074(7)  &0.09(2)   &0.16(6)  &0.08(4)  &&&        \\
         &0.078(11) &0.10(3)   &0.06(2)  &0.010(5) &&&        \\
         &1.09(27)  &1.1(6)    &--       &--       &&&        \\
\end{tabular}
\caption{ }
\label{t40}
\end{table}

\newpage

\begin{table}
\begin{tabular}{|l|l|l|l|l|l|l|l|}
    T   & 3 & 4 & 5 & 6 & 7 & 8 & 9 \\
    R   &          &&&&&&        \\ \hline
3       &301.(1)   &241.(1)    &203.(1)   &191.(1)   &178.(2)  &171.(2)
&164.(3)        \\
        &0.0621(1) &0.0681(3)  &0.0741(4) &0.0750(5) &0.0782(7)&0.079(1)
&0.082(1)        \\
        &0.0511(1) &0.0574((3) &0.0626(4) &0.0682(6) &0.0699(8)&0.073(1)
&0.074(2)        \\
        &8.48(5)   &8.33(8)    &8.3(1)    &8.4(1)    &8.4(2)   &8.5(3)
&8.5(3)        \\  \hline
4       &          &172.4(2)   &135.2(7)  &119.(1)   &110.(1)  &103.(1)
&96.(2)        \\
        &          &0.0804(1)  &0.0911(5) &0.0982(8) &0.103(1) &0.106(2)
&0.111(3)        \\
        &          &0.0661(1)  &0.0744(6) &0.081(1)  &0.088(2) &0.093(2)
&0.092(3)        \\
        &          &8.06(3)    &8.0(1)    &8.17(16)  &8.4(3)   &8.5(3)
&8.3(4)        \\  \hline
5       &          &           &101.1(5)  &85.7(7)   &77.2(9)  &73.(1)
&62.(2)        \\
        &          &           &0.1111(7) &0.123(1)  &0.131(2) &0.139(3)
&0.148(5)        \\
        &          &           &0.0798(6) &0.085(1)  &0.085(2) &0.120(5)
&0.15(7)        \\
        &          &           &8.04(13)  &8.0(2)    &7.9(3)   &9.4(5)
&8.5(7)        \\  \hline
6       &          &           &          &71.2(8)   &62.(1)   &63.(1)
&63.(2)         \\
        &          &           &          &0.141(2)  &0.159(3) &0.162(2)
&0.177(4)        \\
        &          &           &          &0.078(1)  &0.055(1)
&0.0193(3)&0.0212(5)        \\
        &          &           &          &7.9(3)    &7.7(5)   &7.5(3)
&9.0(5)        \\  \hline
7       &          &           &          &          &56.(2)   &58.(2)   &
  \\
        &          &           &          &          &0.172(6) &0.158(4) &
  \\
        &          &           &          &          &0.052(2) &0.0189(5)&
  \\
        &          &           &          &          &7.9(8)   &6.6(4)   &
  \\
\end{tabular}
\caption{ }
\label{t50}
\end{table}

\newpage

\begin{table}
\begin{tabular}{|l|l|l|l|l|l|l|l|}
    T   & 3 & 4 & 5 & 6 & 7 & 8 & 9 \\
    R   &          &          &&&&&        \\ \hline
3       &160.(1)   &123.(1)   &94.(1)    &85.(1)   &79.(2)   &77.(2)   &71.(3)
      \\
        &0.0437(3) &0.0487(4) &0.0517(7) &0.051(1) &0.053(1) &0.051(2)
&0.055(3)        \\
        &0.0414(3) &0.0447(5) &0.048(1)  &0.050(1) &0.050(2) &0.052(2)
&0.051(3)        \\
        &2.50(4)   &2.33(5)   &2.02(7)   &1.9(1)   &1.8(1)   &1.75(16) &1.7(2)
      \\  \hline
4       &108.(1)   &75.4(6)   &51.(1)    &43.(1)   &37.(2)   &33.(2)   &28.(3)
      \\
        &0.465(5)  &0.0570(5) &0.059(1)  &0.065(2) &0.065(4) &0.068(4)
&0.072(8)        \\
        &0.484(7)  &0.0488(5) &0.060(2)  &0.058(2) &0.062(5) &0.060(5)
&0.055(7)        \\
        &2.10(6)   &1.84(4)   &1.54(8)   &1.4(1)   &1.3(2)   &1.2(2)   &1.0(3)
      \\  \hline
5       &82.(1)    &54.9(9)   &29.2(6)   &25.(1)   &17.(1)   &17.5(1.8)&16.(2)
      \\
        &0.0489(8) &0.059(1)  &0.070(2)  &0.071(4) &0.083(8)
&0.094(13)&0.086(16)        \\
        &0.053(1)  &0.062(2)  &0.070(3)  &0.078(8) &0.075(11)&0.052(8)
&0.067(16)        \\
        &1.86(8)   &1.72(9)   &1.22(9)   &1.16(18) &0.90(24) &0.9(3)
&0.81(41)        \\  \hline
6       &71.(1)    &43.(1)    &24.(1)    &16.(1)   &8.(2)    &8.0(2.2)
&6.0(1.7)        \\
        &0.048(1)  &0.066(2)  &0.080(5)  &0.09(1)  &0.12(2)  &0.15(5)  &0.16(3)
       \\
        &0.060(3)  &0.069(3)  &0.088(9)  &0.075(11)&0.035(7) &0.13(9)  &--
  \\
        &1.79(12)  &1.68(12)  &1.4(2)    &0.93(26) &0.59(28) &1.1(8)   &0.7(3)
      \\  \hline
7       &65.(2)    &38.(2)    &19.(1)    &14.(2)   &9.(3)    &9.(3)    &
\\
        &0.050(2)  &0.066(4)  &0.089(9)  &0.081(16)&0.08(3)  &0.11(5)  &
\\
        &0.060(3)  &0.071(6)  &0.094(17) &0.077(22)&0.04(2)  &--       &
\\
        &1.66(15)  &1.5(2)    &1.3(3)    &0.77(37) &--       &--       &
\\  \hline
8       &60.(2)    &33.9(1.5) &17.(2)    &11.(3)   &4.(2)    &         &
\\
        &0.048(2)  &0.072(4)  &0.096(14) &0.10(3)  &0.21(9)  &         &
\\
        &0.063(4)  &0.067(5)  &0.13(5)   &0.07(3)  &0.02(1)  &         &
\\
        &1.59(15)  &1.4(2)    &1.6(5)    &--       &--       &         &
\\  \hline
9       &55.(3)    &32.(3)    &15.(3)    &12.(4)   &         &         &
\\
        &0.051(3)  &0.074(8)  &0.12(3)   &0.10(5)  &         &         &
\\
        &0.061(6)  &0.072(11) &0.14(8)   &--       &         &         &
\\
        &1.5(3)    &1.4(4)    &1.7(9)    &--       &         &         &
\\
\end{tabular}
\caption{ }
\label{t60}
\end{table}

\newpage

\figure{Links involved in link, corner and plaquette integrals.
\label{f10}}

\figure{Surface plots of electric and magnetic components.
For each component two sections are shown, one is on the longitudinal
plane containing the $q\bar q$ pair, another is on the transverse
plane midway between the $q\bar q$ pair, (a) and (b) are for
$\frac{1}{2}E_{\|}^2$, (c) and (d) for $-\frac{1}{2}B_{\|}^2$,
(e) and (f) for $\frac{1}{2}E_{\perp}^2$, (g) and (h) for
 $-\frac{1}{2}B_{\perp}^2$. The data were measured on the
Wilson loop with $T=7$, $R=5$ and $\beta=2.4$.
\label{f20}}

\figure{Surface plots of energy and action, similar to Fig.~\ref{f20},
and obtained from the same data set. (a) and (b) are for the
action density, (c) and (d) for the energy density.
\label{f30}}

\figure{Peak density for $R \times T$ Wilson loop sizes with
$\beta=2.4$ and $R = 3-6$, $T = 3-6$.
For fixed $T$ the points decrease monotonically with $R$.
Triangles: energy density for
$R \times T$ loop; circles: energy density for
$T \times R$ loop; squares: $\|$ components of $E^2$ and $B^2$ only.
\label{f40}}

\figure{Extrapolation of action peak density to infinite Wilson
loop time extent. The extrapolated values are plotted at T=12.
The data were measured on the Wilson loop of the
size $R \times T$ with $\beta=2.4$, for $T = 3-9$ and
$R=3$ (open triangles), $R=4$ (solid triangles), $R=5$ (solid squares),
$R=6$ (open squares) and $R=7$ (solid circles).
\label{f50}}

\figure{Extrapolation of action flux tube width at half maximum to
infinite Wilson loop time extent, labels are similar to Fig.~\ref{f50}.
\label{f60}}

\figure{Extrapolation of exponential tail decay length of action flux
tube to infinite Wilson loop time extent, labels are similar to
Fig.~\ref{f50}.
\label{f70}}

\figure{ Peak values of energy and action density;
solid squares: $\beta=2.5$;
triangles: $\beta=2.4$; open squares: $\beta=2.3$. The two curves are
$1/R$ and $1/R^4$ arbitrarily normalized.
\label{f80}}

\figure{Width at half maximum for energy and action density, labels
are similar to Fig.~\ref{f80}.
\label{f90}}

\figure{ Ratio of integrations of action to energy densities
on center slice, labels are similar to Fig.~\ref{f80}.
\label{f110}}

\figure{Ratio of widths of action to energy profiles on center slice,
labels are similar to Fig. \ref{f80}.\label{f95}}

\figure{Integrations of action density on transverse slices along the
$q\bar q$ axis. The data were measured for $\beta=2.4$ on the Wilson
loops extrapolated for large time for, $R=3$ (open triangles),
$R=4$ (solid triangles), $R=5$ (solid squares) and $R=6$ (open squares).
\label{f150}}

\end{document}